# Influence of anisotropy on the expansion performance of auxetic skin meshing geometries: a finite element study

**Masoumeh Razaghi Pey Ghaleh [1,2,] *, Denis O'Mahoney[1,2], Kevin Moerman[3]**

[1] Mechanical and Industrial Engineering, School of Engineering, Atlantic Technological University, Galway, Ireland

[2] Centre for Mathematical Modelling and Intelligent Systems for Health and Environment (MISHE), School of Engineering, Atlantic Technological University, Sligo, Ireland

[3] Mechanical Engineering, School of Engineering, College of Science and Engineering, University of Galway, Ireland

* Correspondence: masoumeh.razaghipeyghaleh@research.atu.ie

## Abstract

Skin meshing is widely used in severe burn treatment to increase graft coverage. However, the expansion ratios reported by manufacturers differ from clinically observed outcomes. Although skin exhibits anisotropic mechanical behaviour, many computational studies employ isotropic constitutive models, which may contribute to inaccurate predictions of graft expansion. In addition, the mesh pattern can strongly influence expansion performance and material stress, and auxetic meshes, which offer lateral expansion under longitudinal extension, may be particularly promising. This study investigates the combined effects of anisotropy and auxetic mesh geometry on the performance of skin graft expansion. Finite element models of auxetic slit-based geometries, including alternative slits (AS), I-shaped re-entrant slits (IS), and H-shaped re-entrant slits (HS), were developed and subjected to 25% tensile strain. Skin was modelled using the anisotropic Gasser Ogden Holzapfel constitutive formulation, with material parameters fitted to published experimental data for human skin. To study the effect of anisotropy, Langer's line orientations were varied in 15° increments from 0° to 90° relative to the tensile load. For each model variation, area ratio, Poisson's ratio, reaction force, and material stress were evaluated and compared with those obtained using an isotropic Ogden formulation. Results showed that anisotropy strongly influenced mesh expansion behaviour. This suggests that, for auxetic mesh patterns, the direction in which the pattern is created will be clinically important. The effect of anisotropy was observed to be complex and highly dependent on mesh pattern type. The presence of the fibrous reinforcement was observed to enhance or inhibit the auxetic expansion behaviour. In all mesh types studied, the expansion performance is lowest when Langer's lines align with the transverse direction. In other words, skin anisotropy can withhold some of the lateral expansion usually observed for auxetic structures. Conversely, the greatest expansion was typically observed when Langer's lines were close to the applied loading direction. The isotropic models systematically overpredicted stress relative to the anisotropic models, highlighting the importance of accounting for material anisotropy in computational skin-meshing research. Finally, this study found the HS pattern exhibited the strongest auxetic response, producing the highest area ratios and the most negative Poisson ratios. The HS geometry, with Langer's lines parallel to the load direction, achieved approximately 17% and 24% greater expansion than the IS and AS 1$^{st}$ order geometries, respectively. These findings support the use of auxetic structures for skin mesh expansion applications and show that anisotropy is an important factor in both deformation and stress prediction.

## 1. Introduction

Skin meshing is a cost-effective surgical technique commonly used to treat extensive burn injuries. In this procedure, a pattern of regularly spaced incisions is introduced into the skin graft, allowing the tissue to expand when stretched and thereby cover a larger wound area. This approach reduces the amount of donor skin required. The extent of expansion is quantified by the meshing ratio, defined as the ratio of the expanded graft area to its original area prior to stretching [1,2]. However, discrepancies have been found between manufacturer-reported meshing ratios and those achieved in practice [2–5]. Addressing these discrepancies requires an accurate characterisation of the mechanical response of skin undergoing deformations [2].

Research on the mechanical behaviour of human skin has shown that it exhibits a nonlinear, anisotropic, and viscoelastic stress-strain response [2,6,7]. These observations indicate that the mechanical behaviour of skin is influenced by both the magnitude and direction of the applied loads, and that this response is largely governed by the spatial distribution and preferred orientation of collagen fibres within the dermal layer [2,8–10]. Collagen fibres are distributed along Langer's lines, which represent the natural tension directions of skin [2], resulting in direction-dependent stiffness and higher stress levels when the skin is stretched parallel rather than perpendicular to these lines [2,8,9].

Several computational studies of skin meshing have employed isotropic hyperelastic constitutive models, including Yeoh, Mooney–Rivlin, and Ogden formulations [1,2,8,11–15]. These models have been implemented either as single-layer representations (e.g. Yeoh model [1,8,13,16], Mooney–Rivlin [1], Ogden [14,15,17]), in which skin is treated as a homogeneous load-bearing layer, or as bilayer representations, in which the epidermis and dermis are modelled separately (e.g. a bilayer of Yeoh model [11] or a bilayer of Mooney–Rivlin[12]). Although bilayer models account for the layered structure of skin, the skin-meshing simulations reviewed here still appear to retain isotropic constitutive assumptions. In the present study, a dermis-focused representation is adopted because the dermis is the dominant load-bearing layer of skin and exhibits anisotropic behaviour associated with collagen fibre orientation and Langer's lines [9]. The epidermis and hypodermis are therefore not explicitly modelled, avoiding the introduction of additional layer-specific material assumptions while focusing on the primary source of direction-dependent mechanical behaviour. Consequently, the absence of anisotropic constitutive behaviour in current computational skin-meshing studies may limit their ability to capture direction-dependent graft expansion and stress distributions.

Finite element analysis (FEA) is typically employed to simulate the mechanical behaviour of skin meshing, with constitutive parameters calibrated against experimental data [2]. Among nonlinear constitutive models, the isotropic Ogden formulation has shown good agreement with experimental data, whereas the Gasser Ogden Holzapfel (GOH) model provides a more appropriate framework when collagen-induced anisotropy is incorporated [2,9,18]. In the current study, the GOH model was used to represent the anisotropic mechanical behaviour of skin, with material parameters calibrated against experimental data for human back skin reported by Ní Annaidh et al. [9]. In addition, an isotropic Ogden formulation was used as a comparative isotropic model, with parameters calibrated to the averaged anisotropic response.

The objective of this study is to evaluate the influence of auxetic skin meshing geometry and material anisotropy on skin meshing expansion performance. Auxetic patterns, characterised by negative Poisson's ratios, expand laterally under longitudinal stretching, enabling greater overall expansion compared to conventional skin meshing designs [2,19]. This study focuses on slit-based auxetic patterns, as these can generate expansion while minimising the loss of graft material. The investigated auxetic patterns were the alternative slits (AS) of 1$^{st}$ and 2$^{nd}$ order [2,12], I-shaped re-entrant slits (IS) [2,20], and H-shaped re-entrant slits (HS) [2,4].

Finite element models were constructed for each geometry and subjected to 25% uniaxial tensile strain, while Langer's line orientation was varied from 0° to 90° in 15° increments relative to the loading direction. Mesh expansion performance and auxetic behaviour were assessed using area ratio, Poisson's ratio, reaction force, and material stress. These metrics were used to study the influence of anisotropy and to evaluate which geometry produced the highest meshing ratio. Furthermore, the combined effects of graft geometry and constitutive formulation, comparing anisotropic GOH and isotropic Ogden models, were examined to assess their influence on differences in predicted meshing ratios, expansion area, reaction force, and stress distributions.

## 2. Methods

### 2.1. Finite element framework

All geometry processing, finite element model construction, input-file generation, analysis execution, and result extraction were performed using custom MATLAB R2023a code developed within the open-source computational biomechanics library GIBBON [21]. Finite element analyses were carried out using the open-source solver FEBio V4.0 [22]. The GIBBON–FEBio integration was used to generate FEBio input files, assign material definitions and boundary conditions, run the simulations, and import the resulting displacement, reaction force, and stress outputs for post-processing. This provided a fully scripted and reproducible workflow for geometry generation, constitutive modelling, fibre orientation assignment, nonlinear finite element analysis, and output comparison.

### 2.2. Finite element model construction

Four slit-based auxetic skin meshing geometries were investigated: 1$^{st}$ order alternative slits (AS), 2$^{nd}$ order alternative slits (AS), I-shaped re-entrant slits (IS), and H-shaped re-entrant slits (HS) (Figure 1). The geometric parameters controlling each pattern are listed in Table 1 and shown graphically in Figure 1. For each geometry, closed boundary curves were first defined for a quarter segment of the auxetic unit cell. The boundary curves were next resampled to obtain approximately uniform nodal spacing and were then meshed using constrained Delaunay triangulation using GIBBON's *regionTriMesh* function.

Local refinement was applied near sharp features, particularly around slit tips, using GIBBON's *subTriDual* function to improve mesh quality in regions where stress concentrations were expected. The meshed quarter segments of the unit cells were next mirrored and translated. Following this, mirroring the nodes at the boundaries was merged, producing the complete unit-cell meshes shown on the left of each geometry visualisation in Figure 1. The unit cells were then copied and translated, and shared nodes were merged once more to produce the tilings. The pattern parameters were selected so that the resulting models had similar overall in-plane dimensions of approximately 60 × 60 mm², while maintaining a constant slit width of 1 mm across all geometries. This allowed the mechanical response to be comparable across model configurations.

Finally, the planar meshes were extruded in the thickness direction to generate three-dimensional six-noded pentahedral elements. The extruded layer thickness was set to 3 mm, in line with published data on human skin thickness [23].

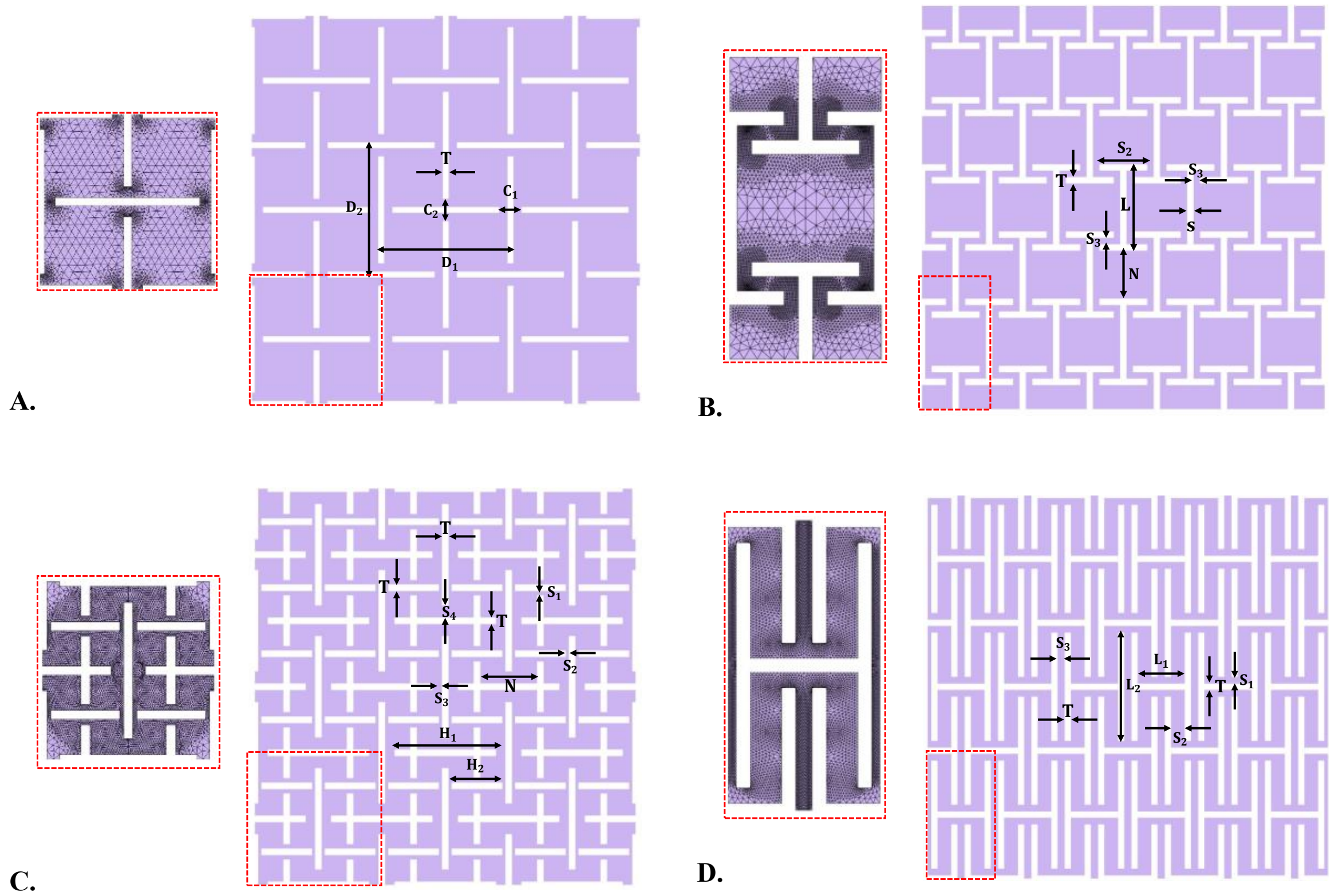

**Figure 1**. Parametric finite element models for auxetic skin graft patterns: A) Alternative slit 1st-order (AS), B) I re-entrant slit (IS), C) Alternative slit 2nd-order (AS), D) H re-entrant slit (HS).

**Table 1.** Parameters for the Auxetic skin meshing pattern geometries

| Auxetic pattern | C1 | C2 | D1 | D2 | T |
|---|---|---|---|---|---|
| AS 1st-order | 3.8 | 3.8 | 22.04 | 22.04 | 1 |

| Auxetic pattern | L | N | S2 | S3 | T |
|---|---|---|---|---|---|
| IS | 9.5 | 7.5 | 7.5 | 1 | 1 |

| Auxetic pattern | H1 | H2 | S1 | S2 | S3 | S4 | N | T |
|---|---|---|---|---|---|---|---|---|
| AS 2nd -order | 15.36 | 7.68 | 0.66 | 0.46 | 1.42 | 1.82 | 8.6 | 1 |

| Auxetic pattern | L1 | L2 | S1 | S2 | S5 | T |
|---|---|---|---|---|---|---|
| HS | 7.6 | 16.56 | 1.08 | 2.25 | 0.99 | 1 |

## 2.3. Constitutive modelling

The mechanical behaviour of skin was modelled using isotropic and anisotropic hyperelastic constitutive formulations to evaluate the influence of the anisotropy on the expansion behaviour of auxetic skin meshes.

The isotropic response was described using the following uncoupled hyperelastic Ogden formulation [24]:

$$\Psi = \sum_{i=1}^{N} \frac{c_i}{m_i^2}\left(\tilde{\lambda}_1^{m_i} + \tilde{\lambda}_2^{m_i} + \tilde{\lambda}_3^{m_i} - 3\right) + \frac{K}{2}(\ln J)^2, \tag{1}$$

where, $c_i$, $m_i$, and $N$ are material parameters of the Ogden model. The deviatoric principal stretches are defined as $\tilde{\lambda}_i = J^{-\frac{1}{3}}\lambda_i$, the eigenvalues of the deviatoric right Cauchy–Green tensor $\tilde{\mathbf{C}}$, and $\lambda_i$ are the principal stretches and $J = \lambda_1\lambda_2\lambda_3$ is the Jacobian, representing the volume ratio. The parameter $K$ The bulk modulus penalises volumetric deformation.

The anisotropic response was represented using an uncoupled Gasser Ogden Holzapfel (GOH) formulation [25]. This model incorporates an isotropic ground matrix of Monney-Rivlin and two dispersed collagen fibre families, allowing the directional mechanical response associated with Langer's lines to be represented. The strain-energy density function was written as:

$$\Psi = \frac{c}{2}(\tilde{I}_1 - 3) + \frac{k_1}{2k_2}\sum_{i=1}^{2}\left(\exp\left(k_2\langle\tilde{E}_i\rangle^2\right) - 1\right) + \frac{K}{2}\left(\frac{J^2-1}{2} - \ln J\right) \tag{2}$$

where $c$ defines the ground matrix stiffness, $k_1$ defines the fibre stiffness, $k_2$ controls the nonlinearity of the fibre response, and $K$ is the bulk modulus. The first deviatoric invariant is defined as $\tilde{I}_1 = tr\tilde{\mathbf{C}}$, where $\tilde{\mathbf{C}}$ is the deviatoric right Cauchy–Green tensor. The Macaulay brackets $\langle\cdot\rangle$ indicate that fibre contributions are active only in tension, so that compressed fibres do not contribute to the strain energy.

The fibre strain measure was $\tilde{E}_i$ is defined as:

$$\tilde{E}_i = \kappa(\tilde{I}_1 - 3) + (1 - 3\kappa)(\tilde{I}_{4i} - 3) \tag{3}$$

where $\tilde{I}_{4i}$ is the squared stretch along the $i^{th}$ fibre family and $\kappa$ is the fibre dispersion parameter. The two fibre families are oriented along directions of $a_{1,2} = \cos\gamma\,\mathbf{i} \pm \sin\gamma\,\mathbf{j}$ within the $\{e_1, e_2\}$ plane, forming angles of $\pm\gamma$ with respect to $e_1$. Fibre dispersion is described by the parameter $\kappa$, where $\kappa = 0$ corresponds to perfectly aligned fibres, $0 < \kappa < \frac{1}{3}$ to dispersed orientations, and $\kappa = \frac{1}{3}$ yields an isotropic fibre distribution. Intermediate $\kappa$ values were defined using a $\pi$ −periodic von Mises fiber distribution to account for fiber orientation periodicity. The parameters, $k_1$, and $k_2$ define the ground matrix stiffness, fibre modulus, and nonlinearity, respectively. In human skin, the fibre angle is commonly taken as $\gamma = \pm 41°$[9].

## 2.4. Constitutive parameter calibration

Experimental data were adopted from uniaxial tensile tests of human back skin reported by Ní Annaidh et al. [9] (Figure 2A). These data include mechanical responses for loading parallel and perpendicular to Langer's lines. Histological analysis in the same study quantified two preferred collagen fibre orientations, from which the mean collagen fibre angle was estimated as $\gamma = 41°$. The fibre dispersion parameter $\kappa$ was determined from the reported fibre orientation distribution using a periodic von Mises distribution ($\kappa = \frac{1}{4}\int_0^{\pi} \rho(\Theta)\sin^3\Theta\, d\Theta$).

To ensure consistency with the present finite element implementation, an independent inverse optimisation procedure was performed using open-source GIBBON [21] in MATLAB, coupled with the Levenberg–Marquardt algorithm, and FEBio V4.0 [22]. This approach is well-suited for nonlinear least-squares problems and provides robust convergence for parameter identification. The model geometry was adapted from Ní Annaidh et al. [9], with dimensions of 6×33×2.25 mm$^3$ (Figure 2A). The geometry was meshed using 8-noded hexahedral elements. Uniaxial boundary conditions were enforced, i.e. the top surface was given a prescribed displacement in the Z-direction, the bottom surface, front surface, and side surfaces were next constrained from moving in the Z, X, and Y direction, respectively. The anisotropic GOH model was implemented as a two-fibre formulation with symmetric collagen fibre families at a fixed mean orientation of $\gamma = \pm 41°$. To preserve the total ground matrix contribution, each fibre-family contribution was assigned half of the matrix stiffness ($c/2$). The remaining GOH parameters of $k_1$, $k_2$, and $\kappa$ were extracted by inverse finite element calibration against experimental data for loading parallel and perpendicular to the collagen fibre direction, with the bulk modulus prescribed as $K = 1000c$ to enforce near-incompressibility. The GOH model reproduced the experimental stress–stretch response with good agreement, with a residual standard deviation of 0.81 MPa relative to the experimental stress range (Figure 2A and Table 2).

An isotropic two-term Ogden model was also calibrated for comparison. For this model, the averaged responses from the parallel and perpendicular experimental datasets were used to represent equivalent isotropic behaviour. The calibrated GOH and Ogden parameter sets were then consistently used in all auxetic skin meshing simulations to assess the influence of anisotropic versus isotropic skin behaviour on area ratio, Poisson's ratio, reaction force, and stress. The optimised material parameters are listed in Table 2. The fitted model responses and the corresponding experimental data are shown in Figure 2.

**Table** 2. Optimised material parameters for the Ogden and GOH models using the Levenberg–Marquardt algorithm

| Constitutive model | $c_1$ [MPa] | $m_1$ | $c_2$ [MPa] | $m_2$ | $K$ | SSE | Std.dev. |
|---|---|---|---|---|---|---|---|
| Ogden | 3.96 | 14.354 | 0.192 | 14.482 | 396 | 0.0123 | $2.346\times10^{-3}$ |
| Constitutive model | $c$ [MPa] | $k_1$ [MPa] | $k_2$ | $\kappa$ | $K$ | SSE | |
| GOH | 1.2099 | 61.790 | $3.4157\times10^{-7}$ | $2.9859\times10^{-1}$ | 1209.9 | 12.375 | $8.0705\times10^{-1}$ |

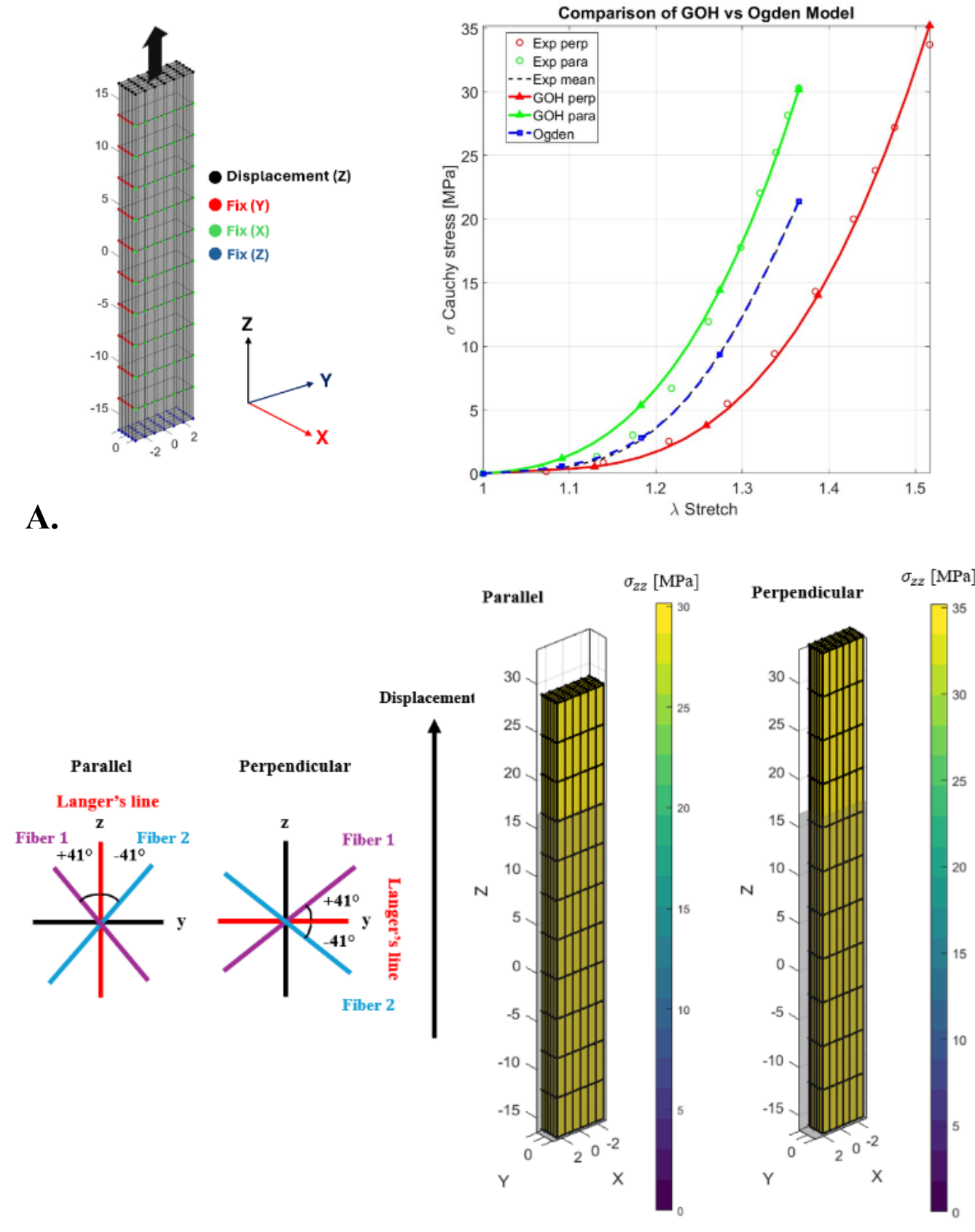

**Figure 2.** Finite element model, constitutive calibration, and stress distribution. **(A)** (Left) Hexahedral FE model of a rectangular specimen under uniaxial tension. (Right) Comparison of GOH and Ogden models with experimental stress–stretch data for loading parallel and perpendicular to Langer's lines. **(B)** (Left) Local material axes and symmetric fibre orientations ($\pm\gamma$) relative to Langer's lines. (Right) Axial Cauchy stress ($\sigma_{zz}$) at maximum stretch.

## 2.5. Boundary conditions and finite element analysis

Following mesh generation, material definitions were assigned using the calibrated parameters listed in Table 2, and FEA was automatically triggered for each model configuration. For each pattern type, eight model variations were studied, i.e. one featuring the isotropic Ogden formulation and seven for the GOH model, whereby the Langer's line angle ranged from 0° to 90° in 15° increments relative to the tensile loading direction (0°, 15°, 30°, 45°, 60°, 75°, and 90°). For the GOH model, fibre directions were computed using Euler-based rotation matrices and assigned elementwise along the structural material axes, as shown in Figures 4-9. Near-incompressibility was enforced using a bulk modulus of $K = 1000c$ for the GOH model and $K = 100c_1$ for the Ogden model.

Boundary conditions (Figure 3) were defined using node-set identification for uniaxial stretch. Each model was subjected to tensile loading in the Y-direction. A prescribed displacement corresponding to an axial stretch of $\lambda_y = 1.25$ (equivalent to 25% tensile strain) was applied to the top surface:

$$\lambda_y = \frac{h}{h_0} \tag{4}$$

where $h_0$ and $h$ are the initial and deformed specimen heights, respectively. The bottom surface was constrained in the Y- direction, while the lateral deformation in the X-direction was permitted to allow expansion or contraction of the auxetic structures. The top and bottom boundaries were constrained in the Z-direction to restrict out-of-plane rigid motion. A single reference node on the bottom boundary was additionally constrained in the X-direction to prevent rigid-body translation. The boundary conditions are shown in Figure 3.

Nonlinear static analysis was performed using full Newton iterations with adaptive time stepping and convergence controls. Solver execution and post-processing were automated through the GIBBON–FEBio workflow, enabling the extraction of displacements, reaction forces, and stresses. Output metrics, including the area ratio, Poisson's ratio, reaction force, and normalised area-weighted 90th percentile stress, were evaluated to characterise the auxetic mechanical response. The computation of these metrics is described in the next section. The fully scripted open-source framework enables reproducible integration of geometry generation, constitutive modelling, fibre orientation, and nonlinear finite-element analysis.

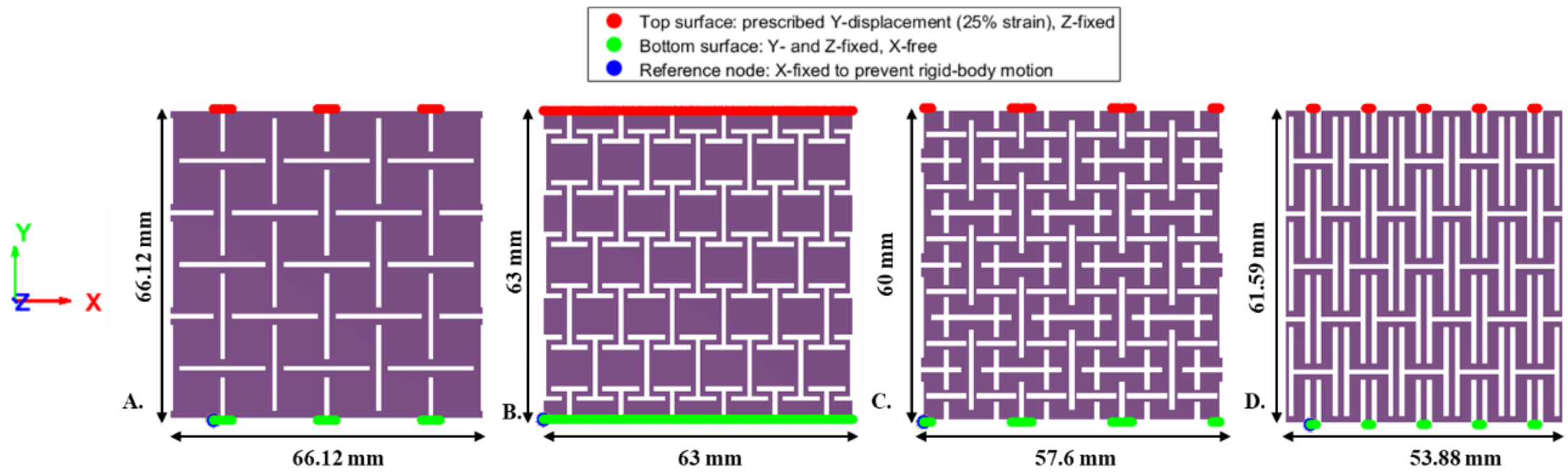

**Figure 3.** Boundary conditions applied to the four FEA models used in the uniaxial tensile simulations: (A) AS 1st order, (B) IS, (C) AS 2nd order, and (D) HS. Red markers indicate the top surface with prescribed Y-displacement, green markers indicate the bottom surface constrained in Y and Z, and the blue marker indicates the reference node fixed in X to prevent rigid-body motion.

## 2.6. Post-processing and model comparison

Post-processing metrics were computed from FEBio output log files imported into MATLAB using GIBBON. The mechanical response of each model was characterised using area ratio, Poisson's ratio, reaction force, and the upper 90th percentile von Mises stress.

First, the apparent transverse stretch $\lambda_x$ was calculated as:

$$\lambda_x = \frac{w}{w_0} \tag{5}$$

where $w_0$ and $w$ are the initial and deformed specimen widths, respectively. The apparent area ratio $J_A$ (a type of global in-plane effective Jacobian) was then computed as:

$$J_A = \frac{wh}{w_0 h_0} = \lambda_x \lambda_y \quad (6)$$

This metric describes the global in-plane expansion of the auxetic graft relative to its initial area.

The apparent Poisson's ratio was calculated as:

$$\nu = -\frac{\ln(\lambda_x)}{\ln(\lambda_y)} \quad (7)$$

where negative values indicate auxetic behaviour, corresponding to lateral expansion under longitudinal extension.

Note that both $J_A$ and v are metrics of the overall auxetic structure rather than the local material behaviour.

The reaction force was calculated by summing the nodal reaction forces in the loading direction over the constrained or prescribed boundary nodes:

$$F_y = \sum_{i=1}^{N} R_{y_i} \quad (8)$$

where $R_{y_i}$ is the reaction force in the Y-direction for the node $i$, and $N$ is the number of boundary condition nodes.

To compare model configurations, the area-normalised von Mises stress was computed. For normalisation, each element's von Mises stress was multiplied by the corresponding element area fraction:

$$q_i = \sigma_{VM,i} \left(\frac{A_i}{A_{tot}}\right), \quad (10)$$

where $\sigma_{\mathrm{VM},i}$ is the von Mises stress of the element $i$, $A_i$ is the corresponding element area, and $A_{\mathrm{tot}}$ is the total model area. The 90$^{\mathrm{th}}$ percentile of the resulting $q_i$ distribution was extracted at each stretch level. This metric was used as a relative stress-contribution measure to compare stress across model configurations.

For each model variation, the area ratio, Poisson's ratio, reaction force, and 90$^{\mathrm{th}}$ percentile of the area-normalised von Mises stress were plotted against applied axial stretch $\lambda_y$. To visualise the influence of anisotropy, the values of each metric at $\lambda_y = 1.25$ were also plotted as a function of Langer's line orientation using polar plots. The results are presented and discussed in Figures 5–8.

# 3. Results

All geometries were evaluated under identical tensile loading conditions up to an axial stretch of $\lambda_y = 1.25$ using calibrated isotropic Ogden and anisotropic GOH constitutive models, as shown in Figure 4. All simulations converged with meshes containing approximately 10,000 elements. Figure 4 summarises the resulting area ratio, Poisson's ratio, reaction force, 90$^{\mathrm{th}}$ percentile of the area-normalised von Mises stress contribution, and lateral displacement for all geometries. Figures 5-8 (E, G, F, H) present the orientation-dependent responses extracted at $\lambda_y = 1.25$, while Figures 5-8 (A, B, C, D) present the full stretch-dependent curves for each geometry, together with polar plots summarising the orientation-dependent response at $\lambda_y = 1.25$.

The results showed that anisotropy strongly influenced mesh expansion behaviour, with the effect complex and dependent on the mesh pattern type. The presence of fibrous reinforcement either enhanced or restricted auxetic expansion depending on geometry and Langer's line orientation. In all mesh types studied, expansion performance was lowest when Langer's lines were closer to the transverse direction. Conversely, the best expansion performance was generally observed when Langer's lines were closer to the applied loading direction.

The area ratio ($J_A$) results, shown in Figures 4 (A) and 5–8 (A, E), demonstrate a clear dependence on both geometry and Langer's line orientation. In all geometries, increasing Langer's line alignment towards the loading direction enhanced transverse expansion, with GOH $75°$–$90°$ generally producing the largest area ratios. The HS geometry achieved the greatest expansion overall, reaching an area ratio of 2.090 under GOH $90°$. This was 23.01% greater than IS under GOH 90°, 37.21% greater than AS 1st order under GOH 75°, and 40.67% greater than AS 2nd order under GOH 75°. Within the HS geometry, the area ratio increased by 10.28% from GOH 0°to GOH 90°, indicating that alignment close to the loading direction enhanced expansion.

The IS geometry exhibited the second-highest expansion response. Its GOH area ratio increased by 5.01% from GOH 0°to GOH 90°, while the isotropic Ogden formulation produced the highest area ratio for this geometry. A similar but weaker orientation-dependent trend was observed for AS 1st order, where the GOH area ratio increased by 2.84% from GOH 0°to GOH 75°. In contrast, AS 2nd order exhibited the lowest area ratios and only minor constitutive sensitivity, with a 3.07% increase from GOH 0° to GOH 75°. These results indicate that HS and IS were more effective at converting axial stretch into global area expansion, whereas AS 2nd order produced the weakest expansion response.

The Poisson's ratio responses, shown in Figures 4 (B) and 5–8 (B, F), showed the same overall geometric ranking as the area ratio results. The HS geometry produced the most negative Poisson's ratio, reaching $-2.120$ under GOH 90°, with a 13.95% increase in magnitude from GOH 0° to GOH 90°. The IS geometry produced the second-most negative response, reaching $-1.4194$ under the isotropic Ogden model and $-1.2955$ under GOH 75°. For IS, the magnitude of the GOH Poisson's ratio increased by 11.87% from GOH 0° to GOH 75°. AS 1st order and AS 2nd order exhibited weaker auxetic responses, with maximum magnitudes of approximately $-0.8909$ and $-0.7725$, respectively. Their GOH Poisson's ratio magnitudes increased by 21.77% and 20.90%, respectively, between the lowest and highest auxetic GOH responses. However, Poisson's ratio alone should not be interpreted as the sole indicator of expansion performance; it should be considered together with area ratio and displacement.

The reaction force responses, shown in Figures 4 (C) and 5–8 (C, G), revealed a strong dependence on the constitutive model and geometry. In AS 1st order, IS, and AS 2nd order, the isotropic Ogden formulation consistently predicted the highest reaction forces. Relative to the Ogden model, the minimum GOH reaction force was 49.69% lower for AS 1st order, 62.03% lower for IS, and 50.72% lower for AS 2nd order. Across the GOH orientations, the reaction force increased by 33.52% for AS 1st order, 66.44% for IS, and 26.35% for AS 2nd order as Langer's line orientation approached the loading direction. Unlike the other geometries, HS exhibited a reduced reaction force with increasing alignment towards the loading direction, decreasing by 10.01% from GOH 0°to GOH 90°. For HS, the minimum GOH reaction force was 45.29% lower than the Ogden value. This indicates that the HS geometry expanded more without a corresponding increase in reaction force.

The stress-related results, shown in Figures 4 (D) and 5–8 (D, H), were evaluated using the 90th percentile of the area-normalised von Mises stress contribution. Within each geometry, the isotropic Ogden formulation produced higher stress values than the GOH formulations. Relative to the Ogden model, the minimum GOH value was 64.54% lower for AS 1st order, 77.95% lower for IS, 56.41% lower for AS 2nd order, and 54.29% lower for HS. These results indicate that the isotropic model systematically predicted higher stresses than the anisotropic model within the same geometry.

Figure 9 shows the final deformed states for the four auxetic geometries in lateral expansion. The displacement contours are in line with the area ratio and Poisson ratio results, with HS exhibiting the greatest lateral displacement, followed by IS, AS 1st order, and AS 2nd order. At $\lambda_y = 1.25$, the maximum lateral displacement in the HS geometry under GOH 90°was 37.65% greater than that of IS under GOH 90°, 82.46% greater than the maximum AS 1st order response, and 3.31 times greater than the maximum AS 2nd order response. Across the GOH simulations, Langer's line alignment closer to the loading direction generally promoted greater transverse deformation, whereas transverse alignment restricted expansion.

Overall, the results demonstrate that auxetic performance was governed by the combined interaction between geometry and constitutive anisotropy. The HS geometry consistently provided the strongest auxetic response, followed by IS, while the AS geometries showed weaker expansion, particularly in the AS 2nd order design. Fibre alignment closer to the loading direction enhanced expansion in most geometries. Conversely, the isotropic Ogden formulation generally improved predictions of reaction force and stress-related responses relative to the anisotropic GOH formulation, suggesting that isotropic assumptions may overestimate the mechanical response of anisotropic skin meshing structures. These results confirm the promising behaviour of auxetic structures for skin meshing expansion applications and show that anisotropy is an important factor in deformation and stress prediction.

## 4. Discussion

This study investigated the combined influence of auxetic geometry and constitutive anisotropy on skin meshing expansion under identical uniaxial loading conditions. The main finding was that material anisotropy influenced the predicted expansion, reaction force, and stress-related response of all auxetic geometries, although the magnitude and direction of this effect depended on the mesh pattern. Fibre alignment closer to the loading direction generally enhances transverse expansion, whereas transverse alignment restricts expansion. Among the geometries studied, the HS pattern produced the highest area ratio and lateral displacement at the prescribed tensile stretch of $\lambda_y = 1.25$, while the isotropic Ogden formulation generally predicted higher reaction forces and stress-related values than the anisotropic GOH formulation. The latter can be explained by the fact that the Ogden formulation parameters were fitted to the mean model response. These results demonstrate the importance of accounting for anisotropy in the computational modelling of auxetic skin grafts.

Previous computational and experimental studies have reported different geometric rankings for auxetic skin meshing. Computational analyses at a 100% strain ratio reported the greatest expansion for AS 1$^{st}$ order geometries, followed by IS and AS 2$^{nd}$ order configurations [12,20], whereas HS showed lower expansion in a separate computational study [4] but favourable expansion under biaxial experimental loading [26]. However, these rankings are not directly comparable with the present results because previous studies used geometry-specific slit dimensions, isotropic constitutive models, different boundary constraints, different loading modes, and larger applied strains. Specifically, Gupta et al. modelled AS 1$^{st}$ and AS 2$^{nd}$ order geometries [12], and HS geometries [4] using bilayer Mooney–Rivlin formulations, geometry-dependent slit widths, and a uniaxial setup in which the left boundary was constrained, and the right boundary was displaced to 100% strain. In contrast, the present study used a consistent slit width across all geometries, single-layer Ogden and anisotropic GOH formulations, permitted lateral deformation, restricted out-of-plane rigid motion, and applied a lower stretch of $\lambda_y = 1.25$. Experimental silicone-based studies are also non-equivalent: AS 1$^{st}$ and AS 2$^{nd}$ order geometries were tested under uniaxial loading in silicone specimens [1,19], whereas IS [20] and AS 1$^{st}$ order, HS, and IS geometries [26] were examined under biaxial loading. Therefore, previous rankings should be considered in the context of qualitative rather than direct validation or contradiction of the present findings, with differences arising from the combined effects of slit geometry, material formulation, boundary conditions, loading mode, and strain level.

The present results also suggest that anisotropy is an important factor in computational models of auxetic skin meshing. For most geometries, alignment of Langer's lines closer to the loading direction increased the area ratio and lateral displacement, whereas alignment closer to the transverse direction reduced expansion. This indicates that collagen fibre orientation can either enhance or restrict the deformation mechanisms of slit-based auxetic structures, depending on the relative orientation between the tissue anisotropy and the imposed loading direction. However, this effect was not uniform across all geometries. The HS geometry showed the strongest expansion at the applied strain level of $\lambda_y = 1.25$, but the broader implication is that anisotropy influences the mechanical response of auxetic skin-meshing patterns and should be considered when predicting graft expansion.

The influence of anisotropy was also reflected in the reaction force and stress-related responses. For AS 1$^{st}$ order, IS, and AS 2$^{nd}$ order geometries, the isotropic Ogden formulation predicted higher reaction forces than the anisotropic GOH formulation. However, this should not be interpreted as a general indication that isotropic models always overpredict structural resistance, since the direction and magnitude of this difference may depend on the material parameters. The HS geometry showed a different response, producing high expansion while maintaining relatively low GOH reaction forces. This suggests that, for this geometry and the loading conditions explored, expansion was achieved through a more compliant deformation mechanism. Nevertheless, this observation should be interpreted within the limits of the present modelling framework and applied strain level, rather than as a general conclusion that one geometry is universally optimal.

The stress-related results further demonstrate the importance of the constitutive formulation. Within each geometry, the isotropic Ogden model produced higher values of the 90$^{th}$ percentile of the area-normalised von Mises stress contribution than the anisotropic GOH model. This difference should be interpreted in relation to the specific material parameterisation used in this study, rather than as evidence that isotropic formulations generally overpredict stress-related responses. More importantly, the results indicate that isotropic and anisotropic constitutive formulations yield distinct stress responses in skin-mesh simulations, underscoring the need to account for anisotropy. However, stress comparisons across different geometries should be treated cautiously, as stress distributions are strongly influenced by the specific geometric configuration, such as the slit arrangement. Therefore, the stress metric is most useful here for relative comparisons within the same geometry and across constitutive formulations.

Although anisotropy influenced all geometries, the degree of sensitivity differed between patterns. The IS geometry showed strong sensitivity to the constitutive formulation in terms of reaction forces and stress-related responses, whereas the AS 2nd order showed comparatively smaller orientation-dependent variation. The HS geometry exhibited the strongest expansion response at the applied strain level of $\lambda_y = 1.25$, but its area ratio varied relatively modestly across GOH orientations compared with the overall difference between geometries. This suggests that geometric topology remained a major factor governing expansion, while anisotropy modified the deformation and stress-related response within each pattern. In other words, anisotropy did not replace the role of geometry but interacted with it.

The present open-source GIBBON–FEBio framework provides a reproducible computational approach for evaluating auxetic skin-meshing patterns. By combining parametric geometry generation, anisotropic constitutive modelling, fibre orientation assignment, and automated finite-element post-processing, the framework enables a systematic investigation of geometry–material interactions relevant to skin graft expansion. The framework is shared openly (https://github.com/MasoumehRazaghi90/skin-auxetic-inverse-analysis), such that other researchers can reproduce and build upon the presented work.

Several limitations should nevertheless be acknowledged. First, the study remains computational and lacks direct experimental validation under identical loading conditions. Material parameters were calibrated using reported human skin data and therefore do not capture patient-specific variability, anatomical location dependence, or multilayer tissue behaviour. The constitutive formulations also assumed homogeneous hyperelastic behaviour and did not include viscoelasticity, damage evolution, tearing, or failure mechanisms, which may become important at larger deformations. In addition, only tensile loading up to 25% was explored. Because auxetic expansion depends strongly on applied stretch, future work could explore higher tensile loads. However, 25% was used here, as it was deemed comparable to what is clinically achievable.

Further limitations relate to the idealised geometry and modelling assumptions. The slit width, pattern dimensions, and model thickness were fixed in the present study to enable controlled comparison between geometries, but other parameter choices may affect the expansion response. As such, alteration of the geometry parameters, as well as evaluation of other auxetic structures, may be explored in future work. In addition, the geometries were also idealised and did not account for manufacturing constraints, surgical handling, graft hydration, biological healing, or tissue heterogeneity. The sensitivity of auxetic graft performance to these parameters may be the topic of future research.

Future work should focus on experimental validation using synthetic or biological skin specimens under controlled uniaxial and biaxial loading conditions. Further studies should also investigate additional strain levels, alternative hyperelastic formulations, viscoelasticity, damage evolution, and multilayer anisotropic tissue models. Parametric optimisation of auxetic geometries within anisotropic modelling frameworks may help identify designs that improve expansion efficiency while controlling stress concentrations, thereby supporting the development of improved skin-meshing strategies for wound treatment applications.

## 5. Conclusion

This study evaluated the combined effects of constitutive anisotropy and auxetic geometry on skin meshing performance under tensile loading using isotropic Ogden and anisotropic Gasser Ogden Holzapfel formulations. The results showed that both mesh geometry and collagen fibre orientation influenced auxetic expansion behaviour, reaction force, and stress-related response.

Among the geometries investigated, the HS geometry produced the greatest expansion, the most negative Poisson's ratios, and the largest lateral displacements under the present loading conditions. Fibre alignment closer to the loading direction generally enhanced auxetic deformation, while reducing reaction force and the stress-related response predicted by the anisotropic GOH model. In contrast, the isotropic Ogden model generally predicted higher reaction forces and stress-related values, suggesting that isotropic assumptions may overestimate the mechanical response of anisotropic skin meshing design.

Overall, the findings demonstrate the importance of incorporating material anisotropy and collagen fibre orientation in computational models of auxetic skin graft expansion. For the tensile strains investigated (up to 25%), the HS geometry showed the strongest expansion response among the configurations studied. Future research should explore other auxetic structure types, as well as other strain levels, and provide a more detailed evaluation of the mechanical behaviour of skin, including the inclusion of skin layers and the representation of viscoelasticity and damage development.

Finally, the computational framework employed here is made available as open source, enabling other researchers to apply and extend the work.

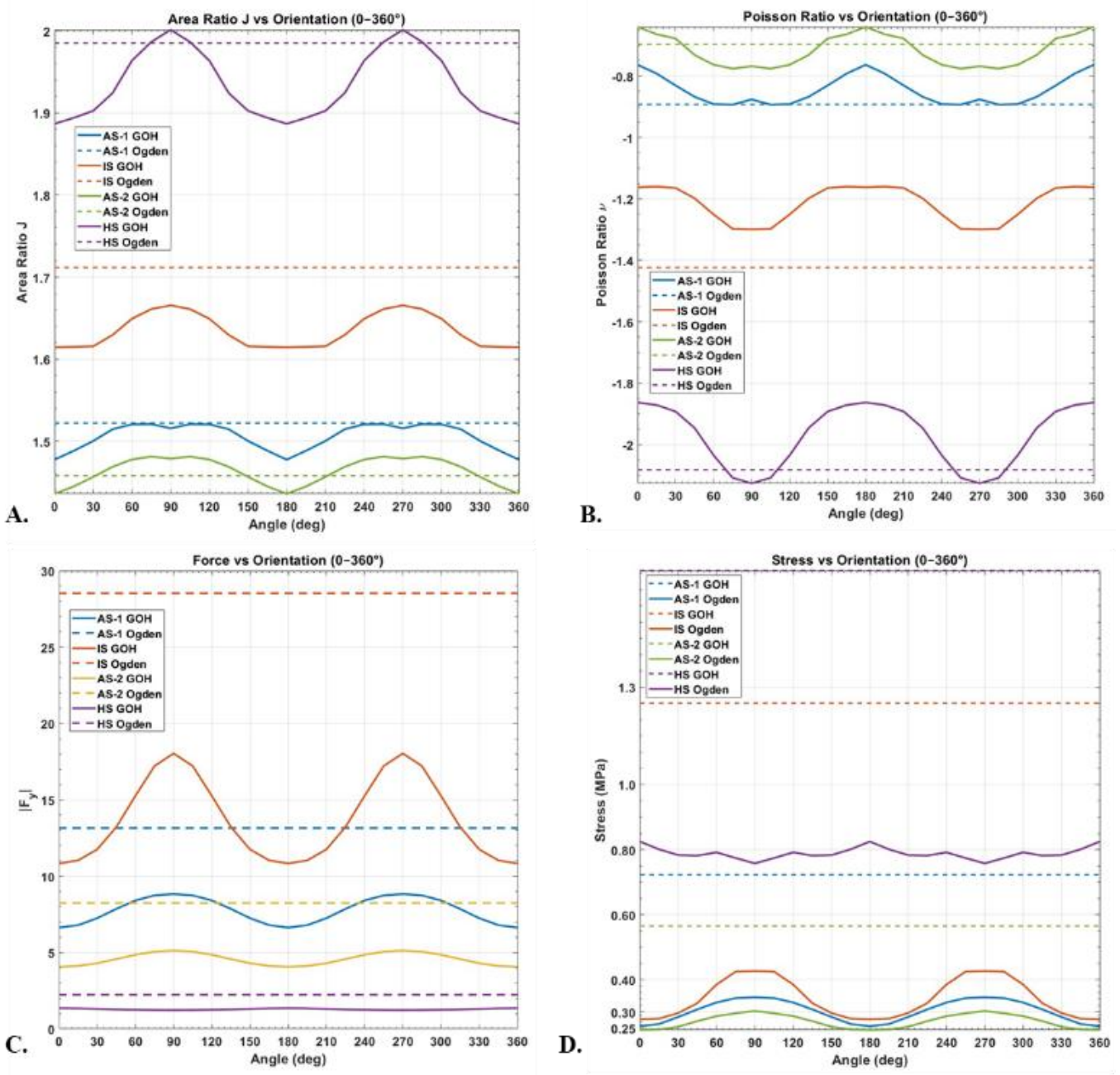

**Figure 4.** Summary of all geometries: (A) maximum area ratio, (B) maximum Poisson's ratio, (C) maximum absolute reaction force, and (D) the maximum area-weighted 90th percentile stress.

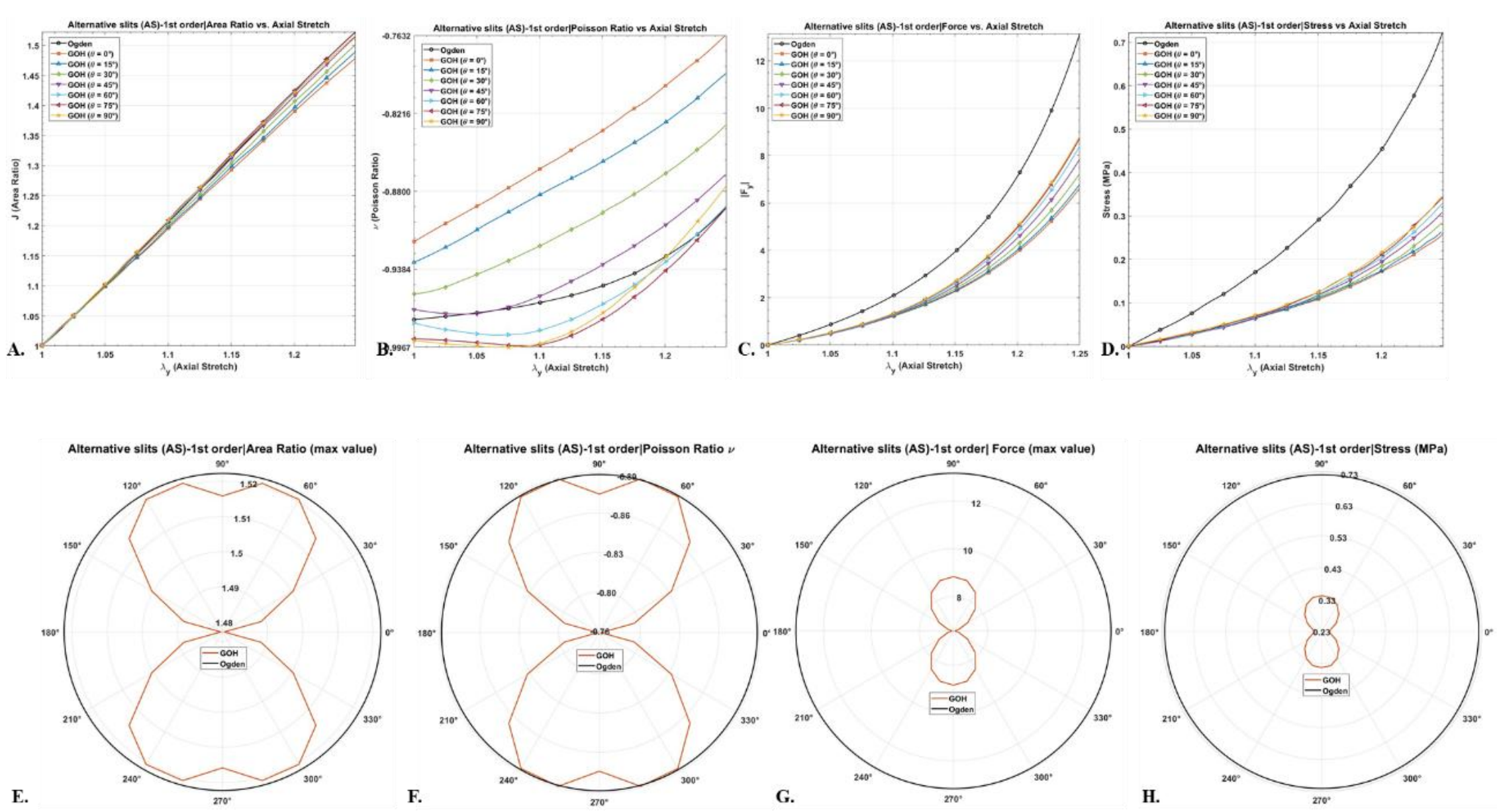

**Figure 5.** Alternative slits (AS)-1st order: (A–D) full-range linear plots of area ratio, Poisson's ratio, absolute reaction force, and 90th percentile stress versus axial stretch. (E–H) maximum values at an axial stretch of $\lambda_y = 1.25$ for area ratio, Poisson's ratio, absolute reaction force, and area-weighted 90th percentile stress.

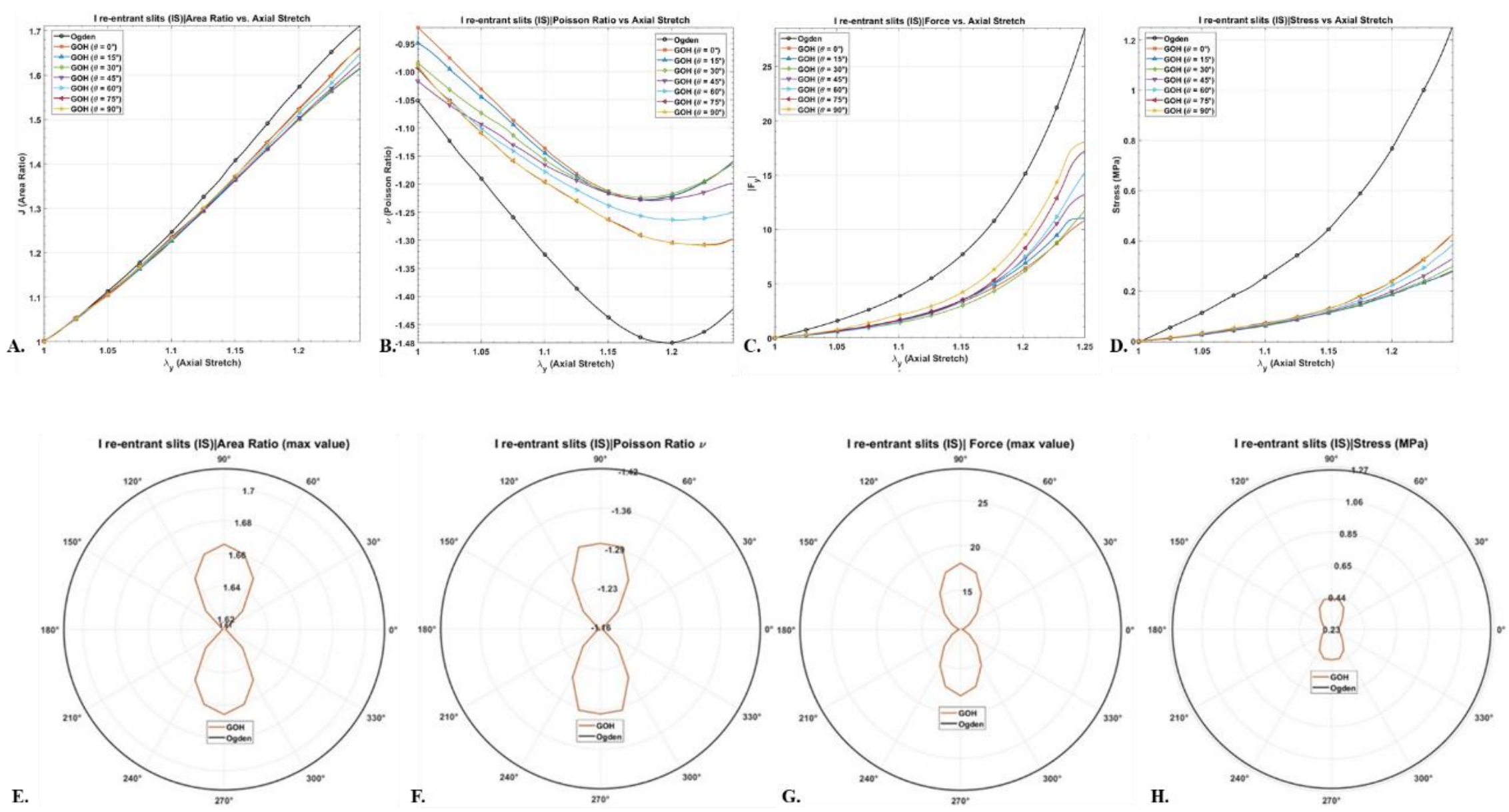

**Figure 6.** I re-entrant slits (IS): (A–D) full-range linear plots of area ratio, Poisson's ratio, absolute reaction force, and 90th percentile stress versus axial stretch. (E–H) maximum values at an axial stretch of $\lambda_y = 1.25$ for area ratio, Poisson's ratio, absolute reaction force, and area-weighted 90th percentile stress.

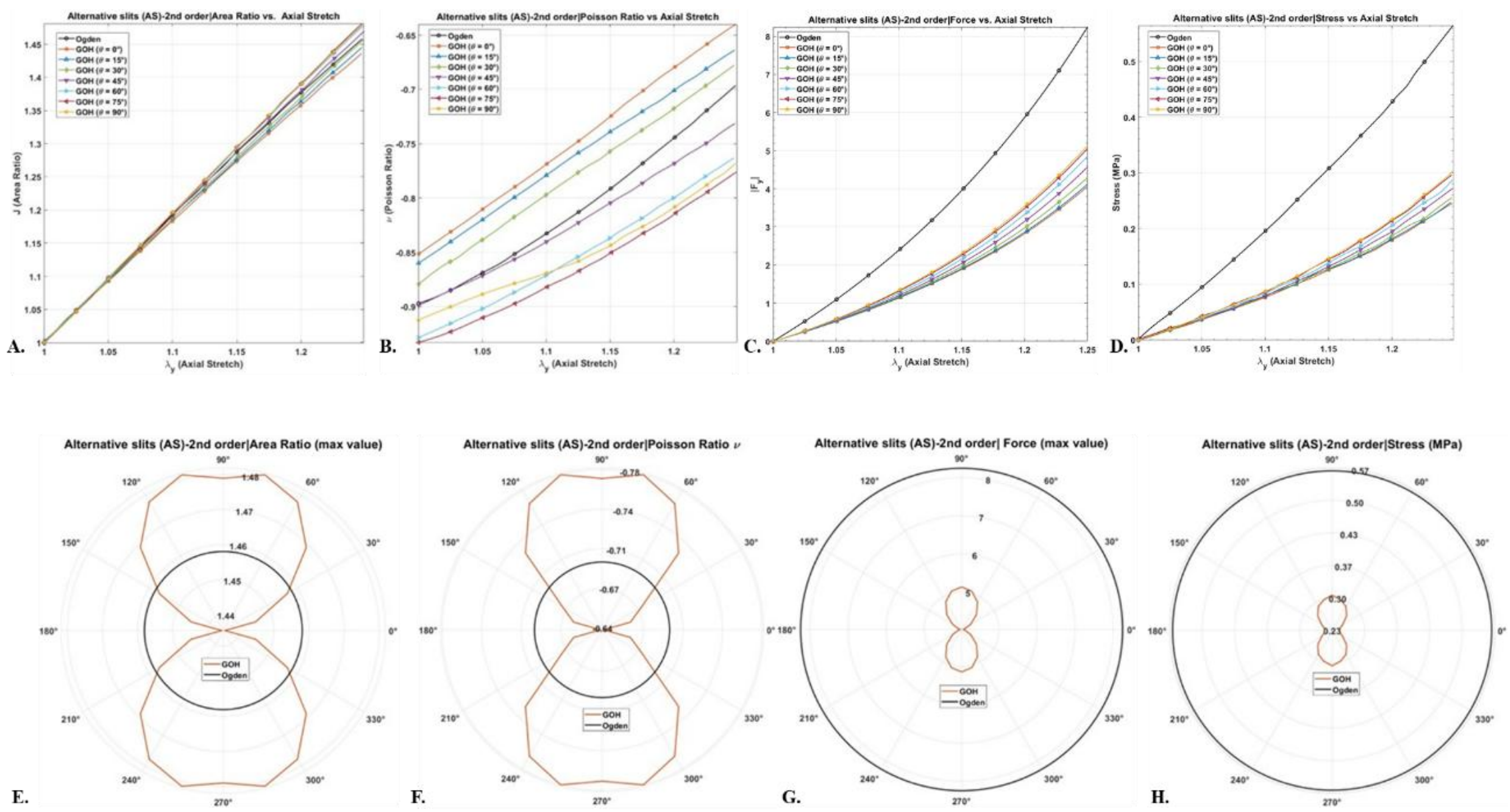

**Figure 7.** Alternative slits (AS)-2nd order: (A–D) full-range linear plots of area ratio, Poisson's ratio, absolute reaction force, and 90th percentile stress versus axial stretch. (E–H) maximum values at an axial stretch of $\lambda_y = 1.25$ for area ratio, Poisson's ratio, absolute reaction force, and area-weighted 90th percentile stress.

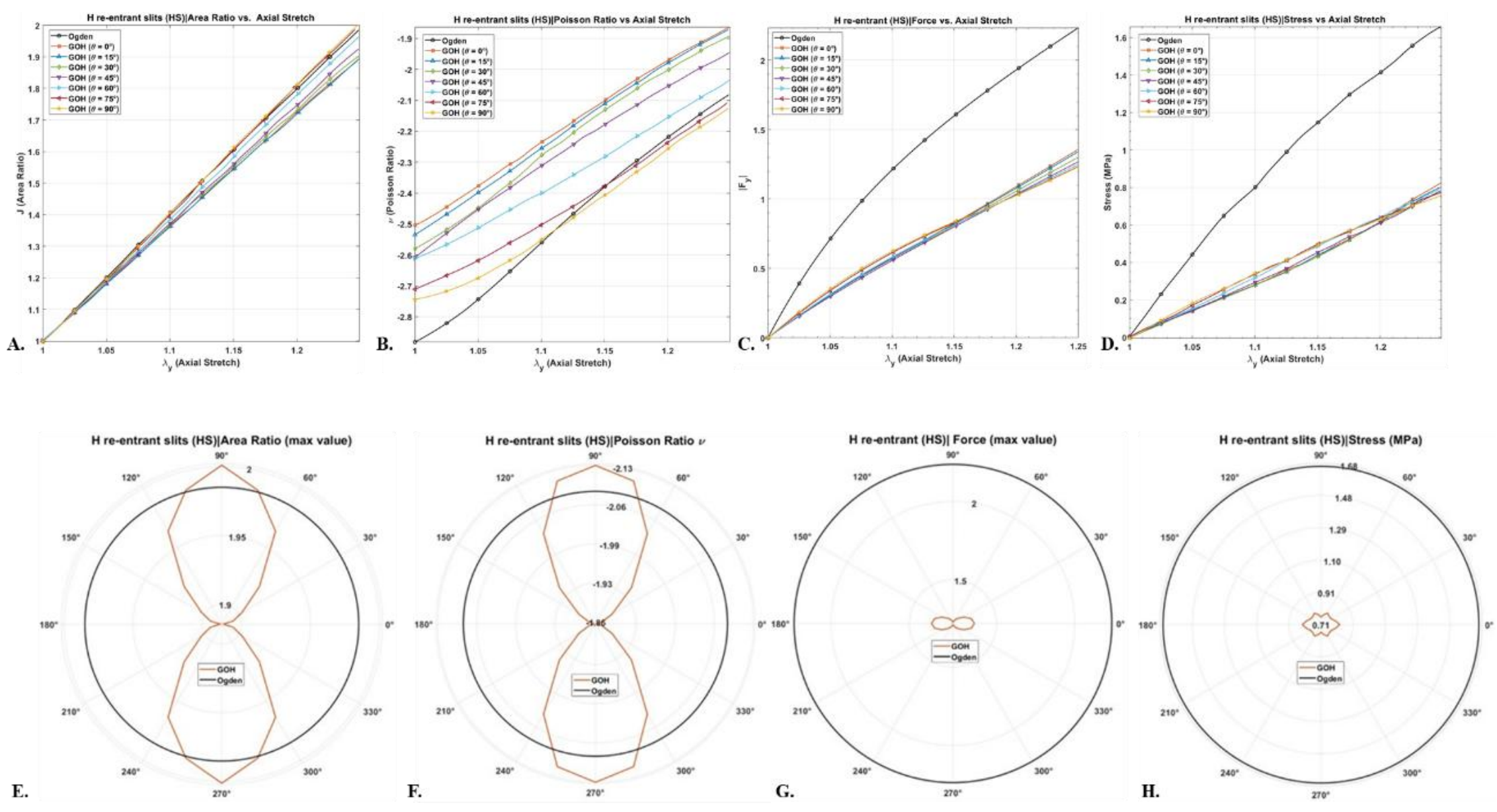

**Figure 8.** H re-entrant slits (HS): (A–D) full-range linear plots of area ratio, Poisson's ratio, absolute reaction force, and 90th percentile stress versus axial stretch. (E–H) maximum values at an axial stretch of $\lambda_y = 1.25$ for area ratio, Poisson's ratio, absolute reaction force, and area-weighted 90th percentile stress.

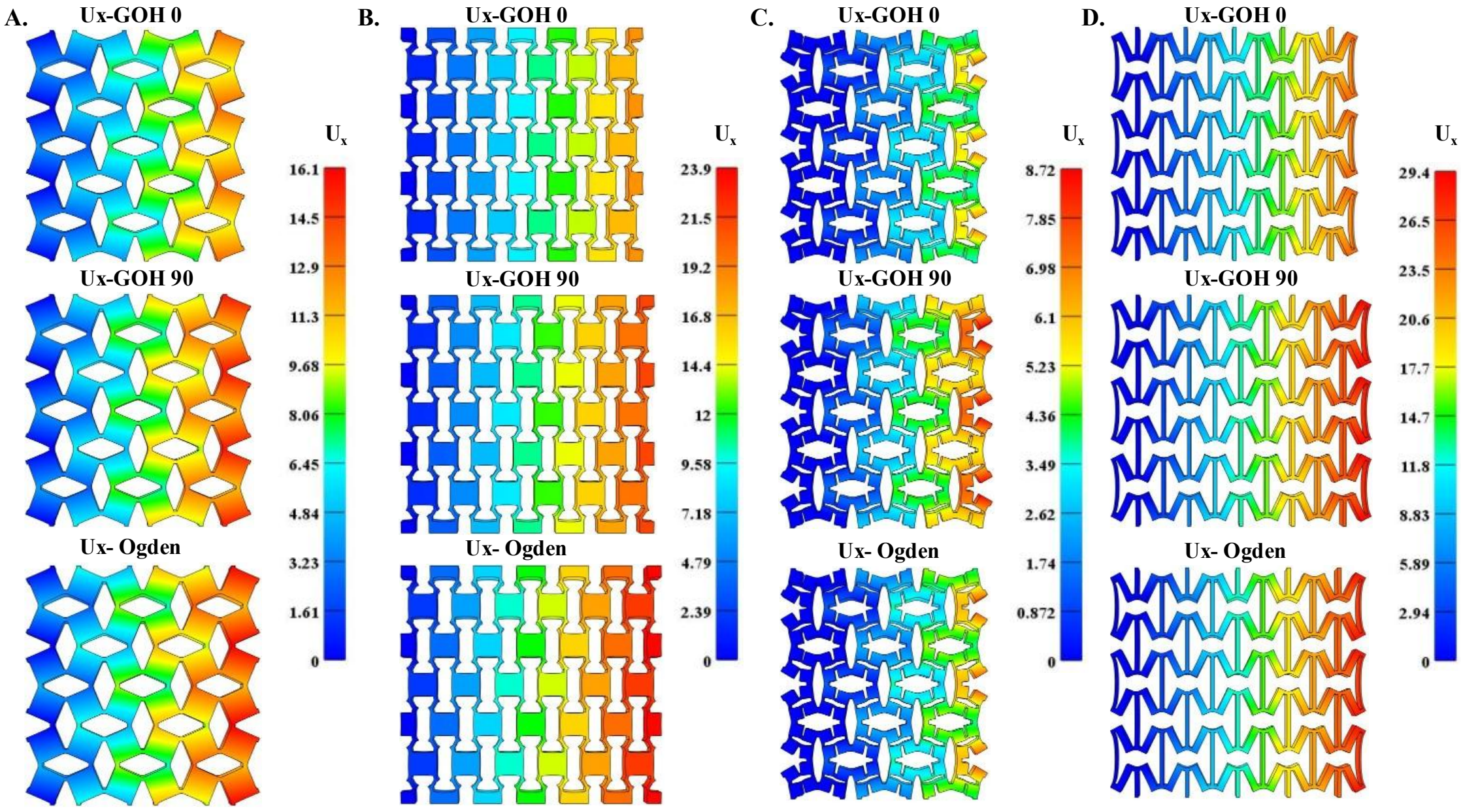

**Figure 9.** Expansion of auxetic geometries along the axial axis: A. AS 1st order, B. IS, C. AS 2nd order, D. HS

## Acknowledgments

This research was supported by the MOCHAS scholarship from Atlantic Technological University.